\documentclass[12pt]{article}
\usepackage{graphicx}
\usepackage{calrsfs}
\usepackage{amstext}
\usepackage{caption}
\usepackage{etoolbox}
\usepackage{makeidx}
\include{epsf}
\usepackage[toc,page]{appendix}
\usepackage{amsfonts}
\usepackage{authblk} 
\usepackage{tcolorbox}
\usepackage{sectsty}
\usepackage{amsmath,amssymb,epsfig}
\usepackage{amscd}
\usepackage{amsthm}
\usepackage{braket}
\usepackage{amsmath}
\usepackage{xcolor}
\usepackage{amssymb}
\usepackage{mathrsfs}
\usepackage{setspace}
\usepackage{url}
\usepackage{dsfont}
\usepackage[applemac]{inputenc}
\usepackage[english]{babel}
\usepackage{enumitem} 
\usepackage{verbatim}
\usepackage{slashed}
\usepackage[]{latexsym}
\usepackage{mathtools}
\usepackage{subcaption}
\usepackage{ mathrsfs }
\usepackage{braket}
\usepackage{caption}
\usepackage{cite}
\usepackage{dsfont}
\usepackage{mathpazo}
\usepackage{hyperref}
\hypersetup{
pdftitle={},%
pdfauthor={},%
pdfsubject={},%
pdfkeywords={},%
colorlinks=true,%
linkcolor=blue,%
citecolor=red,%
linktocpage=true,%
pageanchor=true
}

\newcommand{\be}{\begin{equation}}
\newcommand{\ee}{\end{equation}}
\def\beqa{\begin{eqnarray}}
\def\eeqa{\end{eqnarray}}
\def\bean{\begin{eqnarray*}}
\def\eean{\end{eqnarray*}}

\newcommand{\R}{\mathbb{R}}
\newcommand{\C}{\mathbb{C}}

\theoremstyle{definition}

\newcommand{\eqn}[1]{(\ref{#1})}

\newcommand{\Tr}[1]{\:{\rm Tr}\,#1}

\renewenvironment{thebibliography}[1]
         {\section*{References}\frenchspacing\small
          \begin{list}{[\arabic{enumi}]}
         {\usecounter{enumi}\parsep=2pt\topsep 0pt
         \settowidth{\labelwidth}{[#1]}
         \leftmargin=\labelwidth\advance\leftmargin\labelsep
         \rightmargin=0pt\itemsep=1pt\sloppy}}{\end{list}}

 \numberwithin{equation}{section}

\title{\textbf{On the classical Integrability of Poisson-Lie T-dual WZW models }\vspace{0.5cm}}
\date{}

\author[1]{Francesco Bascone}
\author[1]{Franco Pezzella}
\author[1,2]{Patrizia Vitale}
\affil[ ]{}
\affil[1]{\textit{\footnotesize INFN-Sezione di Napoli, Complesso Universitario di Monte S. Angelo Edificio 6, via Cintia, 80126 Napoli, Italy.}}
\affil[2]{\textit{\footnotesize Dipartimento di Fisica ``E. Pancini'', Universit\`a di Napoli Federico II, Complesso Universitario di Monte S. Angelo Edificio 6, via Cintia, 80126 Napoli, Italy.}}
\affil[ ]{}
\affil[ ]{\footnotesize e-mail: \texttt{francesco.bascone@na.infn.it, franco.pezzella@na.infn.it, patrizia.vitale@na.infn.it}}
\begin{document}
\maketitle
\begin{abstract}
\small We consider the integrability of a two-parameter deformation of the Wess-Zumino-Witten model, previously introduced in 
relation with
Poisson-Lie T-duality.
The resulting family of Poisson-Lie dual models
is shown 
to be integrable by using the Maillet r/s formalism.
 
\end{abstract}

\newpage
\tableofcontents

\section{Introduction}
\label{intro}

Two-dimensional integrable sigma models have become of fundamental importance in many areas of physics, being a class of interacting field theories which can be, in principle, solved exactly (see e.g. \cite{Hoare:2021dix} for a recent review). In particular, much of the recent work in this topic has been carried out for gauge and string theories. In fact, at the worldsheet level, string theories are described by two-dimensional sigma models, for which integrability has an important role for obtaining the exact spectrum of the theories. To give a specific example, note that integrable models arise on the worldsheet description of some string backgrounds which are relevant for the AdS/CFT duality \cite{Maldacena:1997re, Beisert:2010jr}. 

Classical integrable models are characterised by the existence of the so called Lax pair, a couple  of fields  which linearise the equations of motion of the theory, while   satisfying a Poisson algebra which is in general {\it ultralocal}, that is,  the Poisson brackets do not contain  derivatives of the Dirac delta.

However, field theories like the one described by the Wess-Zumino-Witten (WZW) model are {\it non-ultralocal} \cite{Maillet:1985ek}, a condition which introduces discontinuous functions in the Poisson algebra of the monodromy matrices, making it difficult to have a well-defined algebra of conserved charges. In this case, the quantization of the theory is challenging and it is actually still an open problem. 

But in the classical regime, the one in which we are interested in this paper, it is enough to verify that the Poisson algebra of currents could be described in terms of a couple of matrices that fulfill a generalised Yang-Baxter equation. This structure,  introduced by Maillet in \cite{Maillet:1985fn, Maillet:1985ek}, ensures that the global  charges of the theory are actually in involution and conserved. 

An important concept that seems to be related to integrability is that of duality, which plays a key role in theoretical physics, relating apparently different theories. In particular, integrability and Poisson-Lie T-duality have revealed  to be strictly connected \cite{Klimcik:2008eq,Delduc:2013qra,Sfetsos:2013wia}, the latter being a generalisation of the string T-duality of models  with toric-compactified  backgrounds. 
In fact, Poisson-Lie T-duality, introduced in \cite{Klimcik:1995ux, Klimcik:1995dy, Klimcik:1995jn}, represents a genuine generalisation since it is not related to toric compactifications and it does not  require isometries at all for the target background. Specifically, symmetry under Poisson-Lie T-duality transformations is based on the concept of Poisson-Lie dual groups and Drinfel'd doubles, which we shall briefly review in due course. 

In this work we analyse the integrability of a parametric 
family of Poisson-Lie dual models which was  introduced in \cite{Bascone:2020dcn}  by means of  current algebra deformation  techniques \cite{Marotta:2018swj, Bascone:2019tuc, Marotta:2019wfq, Bascone:2020ixw}. 
The resulting current algebra is a two-parameter deformation of the original algebra of the model, 
the semi-direct sum $(\mathfrak{su}(2)\dot\oplus \R^3)(\R)$,  into a fully non-Abelian algebra, following the procedure adopted by  Rajeev and collaborators in \cite{Rajeev:1993rs, Rajeev:1988hq, Rajeev:1988bc}. By choosing a purely imaginary deformation parameter, one can show that the new  current algebra is the Kac-Moody algebra of the Lorentz group, hereafter identified with its universal covering, $SL(2,\mathbb{C})$. This is particularly interesting from the point of view of duality, because the Lorentz group is the  Drinfel'd classical double of the group  $SU(2)$, the dual group being  $SB(2,\mathbb{C})$, the  Borel group of $2 \times 2$ complex upper triangular matrices with unit determinant and real diagonal. The two subgroups $SU(2)$ and $SB(2,\mathbb{C})$ are Poisson-Lie dual groups and the current algebra takes the form of a bialgebra, $\mathfrak{su}(2) \bowtie \mathfrak{sb}(2,\mathbb{C})$\footnote{The symbol $`\bowtie'$ denotes a sum of vector spaces which  entails the adjoint action of each addendum on the other (it is neither direct, nor semidirect).}. 

The Hamiltonian of the model is naturally deformed accordingly. The original target phase space, $T^*SU(2)$, is thus replaced by the group manifold of the Lorentz group, $SL(2, \mathbb{C})$, with a two-parameter family of Hamiltonian models. Since the 
role of the two subgroups is symmetric, both can play the role of target configuration space, obtaining in this way the Poisson-Lie T-duality map as an $O(3,3)$ rotation in  the target phase space, which results in performing an exchange of momenta with configuration space fields. Moreover, a new family of  models is  obtained with configuration space the group $SB(2, \mathbb{C})$, which is dual to the previous one by construction.

Let us remark that this construction  is based on the approach described in  \cite{Rajeev:1993rs}.
However, the dual pairs could also be recovered from the family of models considered in \cite{Sfetsos:2013wia, Georgiou:2016urf, Georgiou:2018hpd}, based on an extension of \cite{Balog:1993es}, which generate the well-known integrable $\lambda$-deformations, as well as generalisations. $\lambda$-models were introduced as integrable deformations of the WZW model, and are related to $\eta$-models, which are instead integrable deformations of the Principal Chiral Model, via Poisson-Lie T-duality and analytic continuation \cite{Hoare:2015gda,Sfetsos:2015nya,Klimcik:2015gba}.

The paper is organized as follows. In section \ref{secalternativeformulation}, we review the alternative canonical formulation of the WZW model on $SU(2)$ with $SL(2,\mathbb{C})$ as target phase space, as introduced in \cite{Bascone:2020dcn} and in particular we describe the two-parameter families of Poisson-Lie dual models. We also briefly review the Drinfel'd double structure of the $SL(2,\C)$ Lie group in terms of its  decomposition in terms of  $SU(2)$ and $SB(2,\C)$.

Section \ref{secIntegrability} contains the original results of the work: we prove the classical integrability of the parametric family of dual models  
by generalising the results of   \cite{Rajeev:1996kk}.

Finally, conclusions are reported in  Section \ref{sectconclusions}.

\section{Alternative canonical formulation of the $SU(2)$ WZW model }
\label{secalternativeformulation}

Let $G$ be a connected Lie group (not necessarily semisimple), whose Lie algebra is denoted by $\mathfrak{g}$ and let us also consider $(\Sigma, h)$ to be a two-dimensional orientable (pseudo) Riemannian manifold.

Let $\{e_i\}$ be a basis in the Lie algebra $\mathfrak{g}$, with $X_i$ denoting  the corresponding left-invariant vector fields.  Any invariant metric $\langle\cdot, \cdot \rangle$ on $\mathfrak{g}$ induces a bi-invariant metric on $G$ defined by $\langle X_i, X_j \rangle=\langle e_i, e_j \rangle$ (we denote the metrics with the same symbol). 
The 
left-invariant dual one-forms $\theta^i$ are defined in the usual way: $\theta^i X_j={\delta^i}_j$. The 
 Maurer-Cartan left-invariant one-form on $G$, given by $\Theta=\theta^i e_i \in \Omega^1(G)\otimes \mathfrak{g}$, shall be needed in order to define the  sigma model on the group manifold.

The field content of the theory is a group-valued field $g: \Sigma \to G$, which is the embedding map of the source space $\Sigma$ into the target  group $G$. 
Thus, the Maurer-Cartan one-form is pulled-back to the source space  $\Sigma$ via $g$, obtaining $g^* \Theta \in \Omega^1(\Sigma) \otimes \mathfrak{g}$. If $G$ can be embedded in $GL(n)$, the latter can be written explicitly as $g^* \Theta=g^{-1}dg$.

The action of the WZW model is thus given by
\begin{equation}
S=\frac{1}{4\lambda^2} \int_{\Sigma} \langle g^{-1}dg \underset{'}{\wedge} \star g^{-1}dg \rangle+\kappa \;S_{WZ},
\end{equation}
where $\star$ is the Hodge star operator, involving the metric of the source space. Hence, the first term contains the dynamics
while the second one is the so-called Wess-Zumino (WZ) term 
\begin{equation}
S_{WZ}=-\frac{1}{24 \pi} \int_{\mathcal{B}} \langle \tilde{g}^{-1}d\tilde{g} \underset{'}{\wedge} d(\tilde{g}^{-1}d\tilde{g}) \rangle,
\end{equation}
which is instead topological. Here $\mathcal{B}$ indicates a $3$-dimensional manifold whose boundary is $\Sigma$, and $\tilde{g}$ is the extension of the field $g$ to  $\mathcal{B}$, i.e. $\tilde{g}|_{\Sigma}=g$. Finally, $\langle \cdot, \cdot \rangle$ denotes a suitable non-degenerate product in the Lie algebra $\mathfrak{g}$.

Although the WZ term entails  the three-manifold $\mathcal{B}$ and depends on the extension, it will finally yield local equations of motion, as
one can easily see by computing the variation of the WZ term. The dependence on the extensions, as well as on the several possibilities of manifolds with the same boundary, $\partial \mathcal{B}=\Sigma$,  is not a problem at the classical level since the variation of the action remains the same up to an irrelevant constant term. As long as suitable normalization conditions are met for the coupling $\kappa$, the dependence on the extension $\tilde{g}$ will give no problems also at the quantum level \cite{Witten:1983tw}.

By parametrising the  two-dimensional source space with local coordinates $(t,x)$ we have 
\begin{equation}
g^{-1}dg=
g^{-1}\partial_t g dt+g^{-1}\partial_x g dx,
\end{equation}
which gives 
\begin{equation}
S=\frac{1}{4 \lambda^2} \int_{\Sigma} d^2 x \, \langle g^{-1}\partial^{\mu}g, g^{-1}\partial_{\mu}g \rangle+\frac{\kappa}{24 \pi}\int_{\mathcal{B}} d^3 y \, \epsilon ^{\alpha \beta \gamma}\langle \tilde{g}^{-1}\partial_{\alpha}\tilde{g},\tilde{g}^{-1}\partial_{\beta}\tilde{g}\,\tilde{g}^{-1}\partial_{\gamma}\tilde{g}\rangle,
\end{equation}
where the indices $\alpha, \beta, \gamma$ run over the three-dimensional $\mathcal{B}$ coordinates and we used Minkowski signature $(1,-1)$ so that $\star dt=dx$ and $\star dx=dt$.


We shall refer to this as the Wess-Zumino-Witten (WZW) model, although the name is often reserved to the WZ model supplemented by  the condition that the parameter $\kappa$ and the coupling constant $\lambda$ be related in such a way to guarantee conformal invariance of the quantum model. 

\subsection{WZW model on $SU(2)$}

Let us now consider the Lie group $SU(2)$ as target space, with $g:\R^{1,1}\rightarrow SU(2)$ the embedding map. The $\mathfrak{su}(2)$ Lie algebra generators are $e_i=\sigma_i /2$, with $\sigma_i$ Pauli matrices, satisfying $[e_i, e_j]=i {\epsilon_{ij}}^k e_k$. The non-degenerate, invariant metric on the algebra is $\langle e_i, e_j \rangle=\frac{1}{2}\delta_{ij}$. 

The WZW model in this case is invariant under the global $SU(2) \times SU(2)$ symmetry. 

The first important observation for our approach is that the equations of motion can be written as a system of two first order partial differential equations:
\begin{equation}
\partial_{t}A-\partial_x J=\frac{\kappa \lambda^2}{4 \pi}\left[A,J \right]
\end{equation}
\begin{equation}
\partial_t J-\partial_x A=-\left[A,J \right]
\end{equation}
with $A=\left(g^{-1}\partial_t g \right)^ie_i=A^i e_i$ and $J=\left(g^{-1}\partial_x g \right)^ie_i=J^i e_i$ the Lie algebra valued currents. This has also to be supplemented with the  boundary condition
\begin{equation}
\lim_{|x| \rightarrow \infty} g(x)=1,
\end{equation}
which makes the solution for $g$ unique. The elements of $G$ that, at fixed time, satisfy this boundary condition form an infinite dimensional Lie group $G(\mathbb{R})$ of smooth maps $\mathbb{R}\ni x \to g(x) \in G$ constant at infinity, equipped with the standard pointwise product. This is a kind of generalisation of the concept of loop group. The corresponding Lie algebra $\mathfrak{g}(\mathbb{R})$ is the space of maps $\mathbb{R} \to \mathfrak{g}$ that are sufficiently fast decreasing at infinity to be square integrable, and we refer to this as a \textit{current algebra}. In particular, we have $\mathfrak{g}(\mathbb{R}) \simeq \mathfrak{g} \otimes C^{\infty}(\mathbb{R})$.


\subsubsection{Hamiltonian description}
We can infer from previous analysis that,  in the Lagrangian formulation of the model, the currents $(J^i, A^i)$ play the role of tangent space coordinates of the target, $TSU(2)(\mathbb{R}) \simeq (SU(2) \times \mathbb{R}^3)(\mathbb{R})$, 
 with $J^i$ the generalised left coordinates of the configuration space and $A^i$ the generalised left coordinates of the fibers. 
This suggests to investigate the Hamiltonian setting, which is indeed very suited  for our purposes. 

The Hamiltonian and (equal-time) Poisson brackets which describe the dynamics of the model are
\begin{equation}\label{hamiltonian1}
H_1=\frac{1}{4\lambda^2}\int_{\mathbb{R}}dx\langle I^2+J^2 \rangle,
\end{equation}
\begin{equation}
\label{poissonbracket1}
\begin{aligned}
{} & \{I_i (x), I_j(y)\}=2\lambda^2\left[{\epsilon_{ij}}^k I_k(x)+\frac{\kappa \lambda^2}{4 \pi} \epsilon_{ijk}J^k(x)\right]\delta(x-y) \\ &
\{I_i (x), J^j(y)\}=2\lambda^2\left[{\epsilon_{ki}}^j J^k(x)\delta(x-y)-\delta_i^j \partial_x\delta(x-y) \right] \\ &
\{J^i(x),J^j(y) \}=0,
\end{aligned}
\end{equation}
where we introduced the canonical momenta $I$ as fiber coordinates of the phase space $\mathcal{P}_1=T^* SU(2)(\mathbb{R})$. $T^* SU(2)$  is a Lie group, the semidirect product $SU(2)\ltimes \R^3$, where $\R^3$ is naturally identified with the  dual Lie algebra,  $ \mathfrak{su}^*(2)$, spanned by the currents $I_i$. Therefore,  $\mathcal{P}_1=  (SU(2) \ltimes  \mathfrak{su}^*(2))(\mathbb{R})$. From the form of the Hamiltonian in \eqref{hamiltonian1} it is clear that the condition on how quickly the currents decay to zero at infinity is necessary for the finitness of energy.

As it can be seen from \eqref{poissonbracket1}, the Poisson algebra is homomorphic to the Lie algebra of $\mathcal{P}_1$, namely the affine algebra $\mathfrak{c}_1=\mathfrak{su}(2)(\mathbb{R})\, \dot{\oplus} \, \mathfrak{a}(\mathbb{R})$ 
with $\mathfrak{a}(\mathbb{R})\simeq \mathfrak{su}^*(2)(\mathbb{R})$  Abelian affine. 
 The phase space $\mathcal{P}_1$ may be then alternatively described by the pair $(J^i, I_i) $ with 
$J^i$ the configuration space coordinates and $I_i$ the fiber coordinates. In terms of the latter, the Hamilton equations of motion read
\begin{equation}\label{eom1}
\partial_{t}I-\partial_x J=\frac{\kappa \lambda^2}{4 \pi}\left[I,J \right]
\end{equation}
\begin{equation}\label{eom2}
\partial_t J-\partial_x I=-\left[I,J \right].
\end{equation}
The model is known to be  Poincar\'e and classically conformally invariant.

The Poisson brackets \eqn{poissonbracket1} 
contain terms proportional to derivatives of the Dirac delta, $\partial_x\delta(x-y)$;  therefore, as announced in the introduction, the model is  non-ultralocal. 
The integrability of these theories becomes then troublesome: it is known that    non-ultralocality introduces discontinuous functions in the Poisson algebra of  monodromy matrices so that the algebra of conserved charges is ill-defined. At the classical level, however, it is still possible to prove integrability, when the Poisson algebra of currents 
may be described in terms of r/s matrices satisfying a deformed  Yang-Baxter equation. This is the so called  Maillet r/s structure \cite{Maillet:1985ek, Maillet:1985fn}, which gives rise to conserved charges in involution. Scope of the paper is to show that such a structure exists for the two-parameter family of WZW models we are going to describe.  


\subsection{Alternative formulation with deformed phase space and current algebra}

The crucial observation on which this work relies is that it is possible to deform the current algebra \eqn{poissonbracket1} to a one-parameter family of fully non-Abelian algebras so that the resulting brackets, together with a one-parameter family of deformed Hamiltonians, lead to an equivalent description of the dynamics, albeit with a different target phase space, which can be identified with the group manifold of $SL(2,\mathbb{C})$\footnote{$SL(2,\C)$ as a real manifold is homeomorphic to $T^*SU(2)$, so that there is no topological obstruction to the alternative picture we are investigating.}. As we shall see, this is a relevant property of the model because  $SL(2,\mathbb{C})$  is the Drinfel'd double of $SU(2)$ in a specific decomposition. In particular, $SL(2,\mathbb{C})$ can be factorized into $SU(2)$ and $SB(2,\mathbb{C})$, which is nothing but the familiar   Iwasawa decomposition, $SL(2,\C) \ni \gamma= k\cdot a \cdot n$, with $k\in SU(2)$, $a\cdot n\in SB(2,\C)$,  the latter  being the Borel subgroup of $2 \times 2$ complex upper triangular matrices with unit determinant and real diagonal elements. 

In \cite{Bascone:2020dcn} we showed how to perform the deformation of the algebra \eqn{poissonbracket1} and we introduced new currents that  render  the bialgebra structure of $\mathfrak{sl}(2,\mathbb{C})$  manifest. From the observation that  the role of the two subalgebras was not symmetric,  we then performed a further deformation, so to obtain a two-parameter family of WZW models, with a perfectly symmetric role of the two subalgebras of currents.

\subsubsection{A glimpse on the Drinfel'd double structure of $SL(2, \C)$}\label{secreviewdrinfeld}

In this section we will briefly review the Drinfel'd double structure of the $SL(2,\C)$ group. 

Let us start by considering the well known fact that $SL(2,\C)$ can be factorized according to $SL(2,\C)= SU(2) \cdot SB(2,\C)$, where $SB(2,\C)$ is the group of $2 \times 2$ complex upper triangular matrices with unit determinant and real diagonal. This means that for any $\gamma \in SL(2,\C)$ one can write the product $\gamma=g\cdot  \ell,  g \in SU(2), \ell \in SB(2,\C)$ (one can also consider the "left" decomposition $\ell \cdot g$). It can be shown that the Lie algebras $\mathfrak{su}(2)$ and $\mathfrak{sb}(2,\mathbb{C})$ are maximally isotropic subalgebras of $\mathfrak{sl}(2,\C)$ with respect to the Killing-Cartan form of the latter, which means that $(\mathfrak{sl}(2,\C), \mathfrak{su}(2), \mathfrak{sb}(2,\C))$ is a so-called Manin triple. For a Lie subalgebra to be maximally isotropic with respect to a non-degenerate (ad)invariant bilinear form it simply requires  that the latter be vanishing on any pair of elements of the algebra, and the maximal property refers to the fact that the set cannot be enlarged while still preserving this property.

In fact, let us consider the real form of the $\mathfrak{sl}(2,\C)$ algebra, represented in terms of rotations and boosts:
\begin{equation}
\begin{aligned}
{\left[e_{i}, e_{j}\right] } &=i \epsilon_{i j}{ }^{k} e_{k} \\
{\left[b_{i}, b_{j}\right] } &=-i \epsilon_{i j}{ }^{k} e_{k} \\
{\left[e_{i}, b_{j}\right] } &=i \epsilon_{i j}{ }^{k} b_{k}.
\end{aligned}
\end{equation}
By using the Cartan-Killing product on $\mathfrak{sl}(2,\C)$ given by $\langle v,w \rangle=2\text{Im}\left[\text{Tr}(vw) \right] $  $ \forall v, w \in \mathfrak{sl}(2,\C)$, it is easy to show that the linear combinations 
\begin{equation}\label{sb2cgen}
\tilde{e}^i=\delta^{ij}\left(b_j+{\epsilon^k}_{j3} e_k\right)
\end{equation}
are dual to the $e_i$ generators of the $\mathfrak{su}(2)$ subalgebra, as $\langle \tilde{e}^i,e_j \rangle={\delta^i}_j$. Moreover, the subspace spanned by $\{\tilde{e}^i \}_{i=1,2,3}$ is maximally isotropic with respect to the same product, being $\langle \tilde{e}^i, \tilde{e}^j \rangle=0$, just like it is for the $\mathfrak{su}(2)$ subalgebra: $\langle e_i, e_j \rangle=0$.

The linear combinations in \eqref{sb2cgen} close a  subalgebra with Lie bracket
\begin{equation}\label{fstruct}
[\tilde{e}^{i}, \tilde{e}^{j}]=i f^{i j}{ }_{k} \tilde{e}^{k},
\end{equation}
where the structure constants are computed to be  $f^{ij}_k=\epsilon^{ij\ell}\epsilon_{\ell 3 k}$. This is the Lie algebra of $SB(2,\C)$, which is solvable. 

All these properties make $(\mathfrak{sl}(2,\C), \mathfrak{su}(2), \mathfrak{sb}(2,\C))$ into a Manin triple.  Moreover, the Lie group $SL(2,\C)=SU(2)\cdot SB(2,\C)$ may be given the structure of  a classical Drinfel'd double (see e.g. \cite{chari} for details), if the two subgroups are endowed with a Poisson structure which is compatible with the group  multiplication. In this case,  the two subgroups are said to be Poisson-Lie dual \cite{chari}.

The concept of Drinfel'd double is at the very foundation of the Poisson-Lie T-duality \cite{Klimcik:1995ux, Klimcik:1995dy, Klimcik:1995jn}, and we direct  the reader to the  existing  literature to keep the article short.

We conclude this brief review by mentioning the fact that a positive-definite Riemannian metric can be defined on $\mathfrak{sl}(2,\C)$ by slightly modifying the other non-degenerate invariant scalar product $2\text{Re}\left[\text{Tr}(vw) \right]$, as follows:
\begin{equation}
 \mathcal{H}_{I J}:=\left(\left(e_{I}, e_{J}\right)\right)=\left(\begin{array}{cc}
\delta_{i j} & -\delta_{i p} \epsilon^{j p 3} \\
-\epsilon^{i p 3} \delta_{p j} & \delta^{i j}+\epsilon^{i l 3} \delta_{\ell k} \epsilon^{j k 3}
\end{array}\right),
\end{equation}
where we introduced the doubled notation $e_I=(e_i, \tilde{e}^i)$.  It is easily verified that   $\left(\left(e_{I}, e_{J}\right)\right)=2\text{Re}\left[\text{Tr}(e_{I}^\dag e_{J})\right]$.  The restriction of this metric on the subalgebra $\mathfrak{sb}(2,\C)$, which will be indicated by $h$, is given by
\begin{equation}
h^{ij}=\delta^{i j}+\epsilon^{i l 3} \delta_{\ell k} \epsilon^{j k 3}.
\end{equation}

\subsubsection{Alternative formulation on $SL(2,\C)$}

On introducing a doubled notation for which $S_I \equiv (K^i, S_i)$, 
let us consider the $\mathfrak{sl}(2,\C)(\R)$ Poisson algebra 
$$\{S_I(x), S_J(y) \}={\left(F_{\tau,\alpha}\right)_{IJ}}^K S_K \delta(x-y)+ \left(C_{\tau,\alpha}\right)_{IJ} \partial_x\delta(x-y)$$ together with the Hamiltonian $$H_{\tau, \alpha}=\lambda^2 \int_{\mathbb{R}} dx \, S_I(x) \left(\mathcal{H}_{\tau, \alpha} \right)^{IJ} S_J(x).$$
By  explicitly writing the algebra in terms of $(K^i, S_i)$, with $K^i$  and $S_i$ respectively spanning the  $\mathfrak{sb}(2,\mathbb{C})(\mathbb{R})$ and   $\mathfrak{su}(2)(\mathbb{R})$ subalgebras, the structure constants ${\left(F_{\tau,\alpha}\right)_{IJ}}^K$ and the central charge $\left(C_{\tau,\alpha}\right)_{IJ}$ are specified as follows
 \beqa
 \{S_i(x), \, S_j(y)\; \}&=&i\alpha{\epsilon_{ij}}^k S_k(x)\delta(x-y)- \alpha^2  C\, \delta_{ij}\partial_x\delta(x-y) \nonumber \\
 \{K^i(x), K^j(y) \}&=& i\tau{f^{ij}}_k K^k(x)\delta(x-y)+\tau^2  C\, h^{ij} \partial_x\delta(x-y) \label{poissonbracket2}  \\ 
 \{S_i(x), K^j(y)\;\}&=&\left[i\alpha{\epsilon_{ki}}^j K^k(x)  +i\tau {f^{jk}}_i S_k(x)  \right] \delta(x-y)  -( C'\delta_i^j - i \bar\tau  C {\epsilon_i}^{j3})  \partial_x\delta(x-y) \nonumber
\eeqa
with  $\alpha, \tau$, purely imaginary parameters. 
Hence, it is possible  to check \cite{Bascone:2020dcn} that the following real linear combinations
\beqa
 S_i(x)&=& \frac{i\alpha}{2\lambda^2(1-\bar\tau^2)(1-\rho^2\bar\tau^2)} \left(I_i(x)- \rho\delta_{ik}J^k(x) \right) \\
 K^i(x)&=&\frac{1}{2\lambda^2 i\alpha (1-\bar\tau^2) (1-\rho^2 \bar\tau^2)}\left[J^k(x)\left({\delta^i}_k-i\bar\tau \rho \epsilon^{i\ell3}\delta_{\ell k} \right)+I_k(x)\left(-\rho\bar\tau^2 \delta^{ik}+i\bar\tau \epsilon^{ik3} \right)\right] \nonumber
\eeqa
lead back to the original dynamics \eqref{eom1}, \eqref{eom2} for the  currents $I$ and $J$, for any $\alpha$ and $\tau$ if the central charges have the form
\begin{equation}
C= \frac{\rho}{\lambda^2 \left(1-\rho^2 \bar \tau^2 \right)^2}, \;\;\; C'=- \frac{1+\rho^2 \bar \tau^2 }{2\lambda^2\left(1-\rho^2\bar \tau^2  \right)^2}
\end{equation}
and
\begin{equation}
\mathcal{H}_{\tau, \alpha}=\begin{pmatrix}
  \frac{1}{(i\alpha)^2}\left[(1+\rho^2 \bar \tau^4)\delta^{ij}-\bar\tau^2(1+\rho^2){\epsilon}^{ip3}\delta_{pq} \epsilon^{jq3} \right] &\left[ i\bar\tau(1+\rho^2){\epsilon}^{ip3}+\rho(1+\bar \tau^2)\delta^{ip}\right]\delta_{pj}\\
  \delta_{ip}\left[- i\bar \tau  (1+\rho^2)\epsilon^{pj3}+\rho(1+\bar \tau^2)\delta^{pj} \right] & (i\alpha)^2(1+\rho^2)\delta_{ij}
 \end{pmatrix}
\end{equation}
where $\rho=\frac{\kappa\lambda^2}{4\pi}$ and $i\bar\tau= i\tau\,i\alpha$.
 
The equations of motion for the fields $K^i, S_i$  have the following form
\beqa
\partial_t S_i&=&-\left[\frac{\rho(1-\bar\tau^2)}{1-\rho^2\bar\tau^2}{\delta_i}^k+i\bar\tau\frac{1-\rho^2}{1-\rho^2\bar\tau^2} {\epsilon_i}^{k3} \right]\partial_x S_k+\frac{1-\rho^2}{1-\rho^2\bar\tau^2 }\delta_{ik} \partial_x K^k\\
 \partial_t K^i &=&-\left[\frac{1-\rho^2 \bar\tau^4}{1-\rho^2\bar\tau^2} {\delta^i}_k-2i\bar\tau\frac{\rho(1-\bar\tau^2)}{1-\rho^2\bar\tau^2}\epsilon^{ik3}-\frac{\bar\tau^2(1-\rho^2)}{1-\rho^2\bar\tau^2}\epsilon^{\ell i 3}{\epsilon_{\ell}}^{k3} \right]\partial_x S_k \nonumber \\ 
 & +& \left[\frac{\rho(1-\bar\tau^2)}{1-\rho^2 \bar\tau^2}{\delta^i}_k-i\bar\tau \frac{1-\rho^2}{1-\rho^2 \bar\tau^2}{\epsilon_k}^{i3} \right]\partial_x K^k+2\lambda^2(1-\bar\tau^2)(1-\rho^2 \bar\tau^2){\epsilon^{ik}}_{\ell} S_k K^{\ell} \nonumber \\ 
 &-& 2i\bar\tau\lambda^2(1-\bar\tau^2)(1-\rho^2 \bar\tau^2){\epsilon_p}^{ik} \epsilon^{p \ell 3} S_k S_{\ell}.
\eeqa
This is a consistent deformation  which leads back to the dynamics of the original model, although the target  phase space is now deformed into $\mathcal{P}_2=SL(2,\mathbb{C})$. Notice that the role of the two subalgebras is made symmetric: the limit $i\tau\rightarrow 0$ reproduces the Kac-Moody algebra $(\mathfrak{su}(2)\dot\oplus \mathfrak{a})(\R)$, whereas the limit $i\alpha\rightarrow 0$ yields $(\mathfrak{sb}(2)\dot\oplus \mathfrak{a})(\R)$. 


\subsubsection{Two--parameter family of Poisson-Lie dual models}
\label{secdualmodels}

Because of the  symmetric role played by the two current algebras described above,  it was shown in \cite{Bascone:2020dcn} that  Poisson-Lie duality could be analysed in an appropriate mathematical framework. Let us first review 
 the Poisson-Lie symmetry of the models described so far. 
 
Poisson-Lie symmetry is essentially a symmetry of the dynamics which is not a symmetry of the geometric tensors characterising the model.  However, the failure from being a symmetry in the standard sense is not arbitrary, but governed by the dual group. In the Hamiltonian approach (see for example  \cite{Klimcik96, stern98, stern99,  Marotta:2019wfq,  Bascone:2019tuc}) this may be summarized as follows. Given the generators of the symmetries of the dynamics, closing the Lie algebra of $G$, say, $V_a\in \mathfrak{X}(M)$, the Lie derivative of  the symplectic form $\omega$ w.r.t. $V_a$ is different form zero, being
\be\label{liepoiform}
\iota_{V_a} \omega = \tilde \theta^a; \;\;\; d  \tilde \theta^a = -\frac{1}{2} {f^a}_{bc}  \tilde \theta^b\wedge  \tilde \theta^c
\ee
and ${f^a}_{bc}$ structure constants of the dual group $G^*$. Namely, $\tilde \theta^a$, the `Hamiltonian' one-forms associated with the generators of symmetries, are not closed but obey the Maurer-Cartan equation of the dual group.

Let us see how this applies to the family of models that we have introduced in the previous subsection. 
By considering $(K^i,S_i) $ as alternative coordinates for the phase space $\mathcal{P}_2$, 
one can associate an Hamiltonian vector field to $K^i$ as usual,  as $X_{K^i}\coloneqq \{\cdot, K^i \}$, which  naturally implies  $\left[X_{K^{i}}, X_{K^{j}}\right]= - X_{\left\{K^{i}, K^{j}\right\}}=- i \tau f_{k}^{i j} X_{K^{k}}$ because of the second relation in \eqref{poissonbracket2}. Therefore, they span the Lie algebra $\mathfrak{sb}(2)$. Moreover, 
\begin{equation}
\omega (X_{K^{j}},X_{K^{k}} )= \{K^j, K^k\}
\end{equation}
Their dual one-forms  $\alpha_j$,   defined by $\alpha_j(X_{K^{k}}) = {\delta_j}^k$ , satisfy the Maurer-Cartan equation 
\begin{equation}\label{MCe}
d\alpha_i(X_{K^{j}}, X_{K^{k}})=-\alpha_{i}([X_{K^{j}}, X_{K^{k}}])=-i\tau {f^{j k}}_i
\end{equation}
Let us consider now \eqn{liepoiform}, with $V_a$ generators of $SU(2)$. We compute
\beqa
d\iota_{V_a} \omega (X_{K^{j}}, X_{K^{k}})&=& X_{K^{j}}(\iota_{V_a} \omega ( X_{K^{k}}))-X_{K^{k}}(\iota_{V_a} \omega ( X_{K^{j}}))-  \iota_{V_a} \omega ([X_{K^{j}}, X_{K^{k}}])\nonumber\\
&= & -X_{K^{j}}(\iota_{X_{K^{k}}} \omega ( V_a))+ X_{K^{k}}(\iota_{X_{K^{j}}} \omega ( V_a))+   i \tau f_{\ell}^{jk}  \omega (V_a, X_{K^{\ell}})
\eeqa
which yields, after some algebra
\be
d\iota_{V_a} \omega (X_{K^{j}}, X_{K^{k}})=- i \tau f_{\ell}^{jk} V_a(K^\ell)
\ee
namely, on comparing with \eqn{MCe}, $\iota_{V_a} \omega$ is proportional to $\alpha_a$ and Eq. \eqn{liepoiform} is satisfied. 


From the structure of the deformed algebra in \eqref{poissonbracket2}, it is immediate to check that the role of the two subalgebras is symmetric, so that the analysis of Poisson-Lie symmetry performed above could be repeated for the $\mathfrak{sb}(2)$ generators. Moreover, the limits $i\tau \to 0$ and $i\alpha \to 0$ reproduce the current algebra structures $(\mathfrak{su}(2) \dot{\oplus} \mathfrak{a})(\mathbb{R})$ and $(\mathfrak{sb}(2,\mathbb{C}) \dot{\oplus} \mathfrak{a})(\mathbb{R})$ respectively. For all other values of the parameters the algebra \eqref{poissonbracket2} is isomorphic to $\mathfrak{sl}(2,\mathbb{C})(\mathbb{R})$. 

It is also possible to check that in the limit $i\tau \to 0$ one not only recovers the current algebra structure of the original $SU(2)$ WZW model, but also the dynamics is recovered. However, in the limit $i\alpha \to 0$ the Hamiltonian becomes singular, meaning that it is not possible to obtain a  WZW model with target space $SB(2,\mathbb{C})$ from this family. This problem is of topological nature and one should not    have expected a different answer, since,  differently from $T^* SU(2)$, the cotangent bundle $T^* SB(2,\mathbb{C})$ is not homeomorphic to $SL(2,\mathbb{C})$. In \cite{Bascone:2020dcn} we find a way to relate a suitably defined $SB(2,\C)$ model to the family described so far and we refer  to that for  details.  

Going back to the review, the key observation is that by virtue of the symmetry, 
one can swap $S$ and $K$ by  an endomorphism of the target phase space $SL(2,\mathbb{C})(\mathbb{R})$, $T: \left. (S,K) \right\vert_x \mapsto \left. (K,S) \right\vert_x$, which is actually an $O(3,3)$ rotation in the phase space. Explicitly relabelling the new generators, one can perform the exchange by writing
\begin{equation}
\tilde{K}(x)=S(x), \quad \tilde{S}(x)=K(x),
\end{equation}
and the resulting  dynamics is given by the following family of dual Hamiltonians and dual Poisson brackets
\begin{equation}\label{dualham}
\tilde{H}_{\tau, \alpha}=\lambda^2 \int_{\mathbb{R}} dx \left[\tilde{K}_i(\mathcal{H}_{\tau, \alpha})^{ij}  \tilde{K}_j+ \tilde{S}^i (\mathcal{H}_{\tau, \alpha})_{ij} \tilde{S}^j+\tilde{K}_i {(\mathcal{H}_{\tau, \alpha})^i}_j\tilde{S}^j + \tilde{S}^i {(\mathcal{H}_{\tau, \alpha})_i}^j \tilde{K}_j\right],
\end{equation}
\begin{equation} \label{dcalgebra}
\begin{aligned}
{} &  \{\tilde{K}_i(x), \tilde{K}_j(y) \}=i\alpha{\epsilon_{ij}}^k \tilde{K}_k(x)\delta(x-y)- \alpha^2 C \delta_{ij}\partial_x\delta(x-y) \\
& \{\tilde{S}^i(x), \tilde{S}^j(y)\} = i\tau{f^{ij}}_k \tilde{S}^k(x)\delta(x-y) +\tau^2 C h^{ij} \partial_x\delta(x-y) \\ 
& \{\tilde{K}_i(x), \tilde{S}^j(y)\}=\left[i\alpha{\epsilon_{ki}}^j \tilde{S}^k(x)  +i\tau {f^{jk}}_i \tilde{K}_k(x)  \right] \delta(x-y) -  ( C'\delta_i^j - i \bar \tau C {\epsilon_i}^{j3})  \partial_x\delta(x-y)
\end{aligned}
\end{equation}
The  new two--parameter family of models has for target configuration space the group manifold of $SB(2,\mathbb{C})$, spanned by the fields $\tilde{K_i}$, while the momenta  of the target phase space are now  $\tilde{S^i}$. Hence, this represents by construction a family of infinite dual models on the Poisson-Lie dual group. We refer to these two as interpolating models, while the original $SU(2)$ WZW and the WZW model with target $SB(2,\mathbb{C})$ are referred to as extremal models.

Although the formulation for the interpolating models has not been obtained from an action principle, nevertheless it is possible to exhibit an action from which it can be derived \cite{Bascone:2020dcn}. Finally, classical conformal invariance can be proven to hold.

\section{Integrability of the two-parameter family}
\label{secIntegrability}

In order to show the integrability of the models, it is convenient to work  with a  different basis for the algebra  $\mathfrak{sl}(2,\mathbb{C})(\mathbb{R})$ where the generators  are given by the complex linear combinations 
\begin{equation}\label{sklr}
\begin{aligned}
{} & S_i=i\alpha\delta_{ij}\left(L^j+R^j \right) \\
& \, K^i=i\tau \left[\left(-i {\delta^i}_j+{\epsilon_j}^{i3} \right) L^j+\left(i{\delta^i}_j+{\epsilon_j}^{i3} \right) R^j \right],
\end{aligned}
\end{equation}
with $L$ and $R$ spanning two commuting copies of the $\mathfrak{su}(2)(\mathbb{R})$ algebra:
\begin{equation}\label{bracketlr}
\begin{aligned}
{} & \{L^i(x), L^j(y) \}={\epsilon^{ij}}_k L^k(x) \delta(x-y)+\gamma_L \delta^{ij}\partial_x\delta(x-y) \\ & 
\{R^i(x), R^j(y) \}= {\epsilon^{ij}}_k R^k(x) \delta(x-y)-\gamma_R \delta^{ij}\partial_x\delta(x-y) \\ & 
\{L^i(x), R^j (y) \}=0,
\end{aligned}
\end{equation}
and  $\gamma_L, \gamma_R$ the following  central charges
\begin{equation}
\gamma_L=\frac{1}{4\lambda^2 \bar{\tau}(1-\rho \bar{\tau})^2}, \quad \gamma_R=\frac{1}{4\lambda^2\bar{\tau}(1+\rho\bar{\tau})^2}.
\end{equation}
The latter are related to the previous ones by  $\gamma_L-\gamma_R=C$ and $-\bar{\tau}(\gamma_L+\gamma_R)=C'$.
In this basis the equations of motion acquire a   simpler form,
\begin{equation}\label{eomlr}
\begin{aligned}
{} & \partial_t L^i+\frac{1+\bar\tau^2}{2\bar\tau}\partial_x L^i+\frac{(1-\bar\tau^2)(1+\rho\bar\tau)}{2\bar\tau(1-\rho\bar\tau)}\partial_x R^i-2\lambda^2(1-\bar\tau^2)(1-\rho^2 \bar\tau^2){\epsilon^i}_{jk}L^jR^k=0 \\ &
\partial_t R^i-\frac{(1-\bar\tau^2)(1-\rho\bar\tau)}{2\bar\tau(1+\rho\bar\tau)}\partial_x L^i-\frac{1+\bar\tau^2}{2\bar\tau} \partial_x R^i+2\lambda^2(1-\bar\tau^2)(1-\rho^2 \bar\tau^2){\epsilon^i}_{jk}L^jR^k=0.
\end{aligned}
\end{equation}
The new basis  allows to recognise that the family of models admits an associated  Lax connection  $\mathcal{L}(\zeta)$, which is conserved and flat,  with $\zeta$ the spectral parameter. In fact, it is possible to show that it admits an associated auxiliary linear system
\begin{equation}\label{associatedlinear}
d\psi=\mathcal{L}\psi,
\end{equation}
for some function $\psi$, with $\mathcal{L}=U dt+V dx$ the matrix valued connection one-form, such that the equations of motion of the models \eqref{eomlr} can be obtained from the flatness condition of the Lax connection $d\mathcal{L}+\mathcal{L}\wedge \mathcal{L}=0$ \footnote{This condition follows from the consistency of \eqref{associatedlinear}, by acting with the exterior derivative on both sides of the equation.}, or, explicitly:
\begin{equation}
\partial_t V(t,x;\zeta)-\partial_x U(t,x;\zeta) +[V(t,x;\zeta), U(t,x;\zeta)]=0,
\end{equation}
for arbitrary $\zeta$.
It is possible to check that such a connection is given by:
\begin{equation}\label{laxlr}
\begin{aligned}
{} & V=a(\zeta)L(t,x)+b(\zeta) R(t,x) \\
& U=c(\zeta)L(t,x)+d(\zeta) R(t,x),
\end{aligned}
\end{equation}
with coefficients
\begin{equation}
\begin{aligned}
{} & a(\zeta)= \lambda^2(1-\bar\tau^{2})(1-\rho \bar\tau)\left[\frac{1-\rho}{1-\zeta}(1-\bar\tau)-\frac{1+\rho}{1+\zeta}(1+\bar\tau)\right]\\
& b(\zeta)= \lambda^2(1-\bar\tau^{2})(1+\rho \bar\tau)\left[\frac{1-\rho}{1-\zeta}(1+\bar\tau)-\frac{1+\rho}{1+\zeta}(1-\bar\tau)\right] \\
& c(\zeta)= \lambda^2(1-\bar\tau^{2})(1-\rho \bar\tau)\left[\frac{1-\rho}{1-\zeta}(1-\bar\tau)+\frac{1+\rho}{1+\zeta}(1+\bar\tau)\right] \\
& d(\zeta) = \lambda^2(1-\bar\tau^{2})(1+\rho \bar\tau)\left[\frac{1-\rho}{1-\zeta}(1+\bar\tau)+\frac{1+\rho}{1+\zeta}(1-\bar\tau)\right].
\end{aligned}
\end{equation}
Because of its dependence on the currents $L$ and $R$, the Lax connection is valued in the Lie algebra $\mathfrak{su}(2)$ or in $\mathfrak{sl}(2,\mathbb{C})$, depending on the coefficients to be real or complex, also according to the spectral parameter domain.

Once a Lax connection is found, one can just follow the usual procedure to construct the monodromy matrix, which is given by \footnote{This is actually the infinite volume limit of the monodromy matrix, the latter being properly defined as $\mathcal{M}(t,x,y ; \zeta)=\operatorname{\hat{P}\operatorname {exp}}\left(-\int_x^y d x' V(t, x'; \zeta)\right)$.}
\begin{equation}
\mathcal{M}(t ; \zeta)=\operatorname{\hat{P}\operatorname {exp}}\left(-\int_{\mathbb{R}} d x V(t, x ; \zeta)\right),
\end{equation}
where $\operatorname{\hat{P}\operatorname {exp}} $ denotes a path ordered exponential (greater $x$ to the left).

Because of the flatness condition of the Lax connection, the infinite volume limit of the monodromy matrix is conserved:
\begin{equation} \label{consmon}
\partial_t \mathcal{M}(t;\zeta)=\mathcal{M}(t;\zeta)\left. U(\zeta)  \right\vert_{x \to -\infty}-\mathcal{M}(t;\zeta)\left. V(\zeta) \right\vert_{x \to +\infty}=0
\end{equation}
thanks to the decaying boundary conditions of $L$ and $R$ spanning a current algebra, i.e. $\lim_{|x| \to \infty}L(x)=\lim_{|x| \to \infty}R(x)=0$.

By definition, $\mathcal{M}$, as a function of $\zeta$, is an element of the loop group of $SL(2,\mathbb{C})$ or $SU(2)$ depending on the domain of the spectral parameter. In fact, since $\lim_{|\zeta| \to \infty}\mathcal{L}_x=0$,  from the definition $\lim_{|\zeta| \to \infty}\mathcal{M}=\mathbb{1}$ and by noting that $V^{\dagger}(t,x;\zeta)=-V(t,x;\zeta^*)$ one has that $\mathcal{M}^{\dagger}(\zeta)\mathcal{M}(\zeta)=\mathbb{1}$. The latter implies that if $\zeta \in \mathbb{C}$ then $\mathcal{M}(\zeta) \in LSL(2,\mathbb{C})$ or if $\zeta \in \mathbb{R}$ it is $\mathcal{M}(\zeta) \in LSU(2)$.  $LG$ denotes the loop group of the Lie group $G$, namely    $LG=\left\{\gamma: S^{1} \rightarrow G \mid \gamma \in C^0\left(S^{1}\right)\right\}$, equipped with the standard topology of continuous maps.

\subsection{Maillet r/s structure}

Given that the Poisson algebra \eqn{bracketlr} is non-ultralocal, we shall prove the integrability by showing that it is possible to express the Poisson brackets of the Lax connection in terms of a couple of matrices, $r$ and $s$ in the notation of Maillet, which satisfy the appropriate deformation of a Yang-Baxter equation, as in \cite{Rajeev:1996kk}. 

By using  Eqs. \eqref{laxlr} and the Poisson algebra of the $L$, $R$ currents, Eqs.   \eqref{bracketlr} one  finds, after some algebra, 
\begin{equation}\label{VV}
\begin{aligned}
\left\{V^i(x, \zeta), V^j(y, \xi)\right\} {} & ={\epsilon^{ij}}_k\Big[\Gamma(\zeta, \xi) V^k(x, \zeta)+\Gamma(\xi,\zeta) V^k(y, \xi)\Big] \delta(x-y) \\ & +\Delta(\zeta,\xi) \delta^{ij} \partial_x\delta(x-y),
\end{aligned}
\end{equation}
with
\begin{equation}\label{gamma}
\Gamma(\zeta,\xi)=a(\xi)b(\xi)\frac{a(\zeta)-b(\zeta)}{a(\zeta)b(\xi)-a(\xi)b(\zeta)},
\end{equation}
\begin{equation}
\Delta(\zeta,\xi)=b(\zeta)b(\xi)\gamma_R-a(\zeta)a(\xi)\gamma_L.
\end{equation}
In order to use a matrix notation for the algebra, we pose
\be
\hat{\Gamma}(\zeta,\xi)=\Gamma(\zeta,\xi)E, \;\;\;  \hat{\Delta}=\Delta(\zeta,\xi) E
\ee
 with $E=\delta^{ij}e_i \otimes e_j \in \mathfrak{su}(2)\otimes \mathfrak{su}(2)$. We thus define $r(\zeta, \xi)$ and $s(\zeta, \xi)$ respectively as the 
 skew-symmetric and  symmetric part  of $\hat\Gamma$,
 \begin{equation}
\begin{aligned}
{} & r(\zeta,\xi)=\frac{1}{2}\Big(\hat\Gamma(\zeta,\xi)-\hat\Gamma(\xi,\zeta) \Big) \\
& s(\zeta,\xi)=\frac{1}{2}\Big(\hat\Gamma(\zeta,\xi)+\hat\Gamma(\xi,\zeta),
\end{aligned}
\end{equation}
and introduce the matrix notation 
$V_1=V\otimes \mathbb{1}$ and $V_2=\mathbb{1} \otimes V$, with $V=V^i e_i$. Then,  the Poisson algebra \eqn{VV} is rewritten as 
\begin{equation}\label{laxpoisson}
\begin{aligned}
\left\{V_1(x, \zeta), V_2(y, \xi)\right\} {} & =\left[r(\zeta,\xi),V_1(x, \zeta)+V_2(y, \xi) \right]\delta(x-y)\\ &-\left[s(\zeta,\xi),V_1(x,\zeta)-V_(y,\xi) \right] \delta(x-y)\\ & -2s(\zeta,\xi) \partial_x\delta(x-y),
\end{aligned}
\end{equation}
where the $r$ and $s$ matrices explicitly read
\begin{equation}\label{rmat}
\begin{aligned}
r &=2 \lambda^2 \frac{\left(1-\bar\tau^{2}\right)}{\zeta - \xi}\left[\frac{\zeta^{2}\left(1-\rho^{2} \bar\tau^{2}\right)-2 \zeta \rho\left(1-\bar\tau^{2}\right)+\rho^{2}-\bar\tau^{2}}{\zeta^{2}-1}\right.\\
&\left.+\frac{\xi^{2}\left(1-\rho^{2} \bar\tau^{2}\right)-2 \xi \rho\left(1-\bar\tau^{2}\right)+\rho^{2}-\bar\tau^{2}}{\xi^{2}-1}\right] E,
\end{aligned}
\end{equation}
\begin{equation}\label{smat}
\begin{aligned}
s &=2 \lambda^2 \frac{\left(1-\bar\tau^{2}\right)}{\zeta - \xi}\left[\frac{\zeta^{2}\left(1-\rho^{2} \bar\tau^{2}\right)-2 \zeta \rho\left(1-\bar\tau^{2}\right)+\rho^{2}-\bar\tau^{2}}{\zeta^{2}-1}\right.\\
&\left. -\frac{\xi^{2}\left(1-\rho^{2} \bar\tau^{2}\right)-2 \xi \rho\left(1-\bar\tau^{2}\right)+\rho^{2}-\bar\tau^{2}}{\xi^{2}-1}\right] E.
\end{aligned}
\end{equation}
Note that the Poisson brackets in \eqref{laxpoisson} are taken at different space points but also at different values of the spectral parameter. Interestingly, the Jacobi identity for the brackets \eqn{laxpoisson}, yields an equation for the $r$ and $s$ matrices
\begin{equation}\label{modiYB}
\begin{aligned}
{} & [(r-s)_{12}(\zeta_1,\zeta_2),(r+s)_{13}(\zeta_1,\zeta_3)]+[(r+s)_{12}(\zeta_1,\zeta_2),(r+s)_{23}(\zeta_2,\zeta_3)]\\ & +[(r+s)_{13}(\zeta_1,\zeta_3),(r+s)_{23}(\zeta_2,\zeta_3)]=0
\end{aligned}
\end{equation}
that is verified by construction, since the original current algebra is already known to satisfy Jacobi identity. The latter becomes the standard Yang Baxter equation for the matrix $r$ when $s$ is zero.

As it was already anticipated in Sec. \ref{secalternativeformulation}, the Poisson brackets between the spatial components of the Lax connection contain central terms proportional to $\partial_x\delta(x-y)$, being therefore referred to as non-ultralocal. In general, this kind of models may exhibit a space-time dependence for $r$ and $s$ matrices, the general form of the algebra  thus being
\begin{equation}
\begin{gathered}
\left\{V_1(x, \zeta), V_2(y, \xi)\right\}=\Big(\partial_x r(x, \zeta,\xi)+\left[r(x, \zeta,\xi),V_1(x, \zeta)+V_2(x, \xi)\right] \\
-\left[s(x, \zeta,\xi), V_1(x,\zeta)-V_2(x, \xi)\right]\Big) \delta(x-y)-\Big(s(x, \zeta,\xi)+s(y, \zeta,\xi)\Big) \partial_x\delta(x-y).
\end{gathered}
\end{equation}
In principle there could be higher derivatives of the delta function; 
 in our case, however, the r and s matrices are non-dynamical.
 
Note that the algebra \eqn{laxpoisson} is well defined for every value of the parameters (we recall that $\bar\tau= i \tau\alpha$), including the limits $i\tau \to 0$ and $i\alpha \to 0$, which are singular for the generators $L$ and $R$, as they cease to be independent functions of $S$ and $K$. 

Once the Poisson algebra of the Lax connection has been put in the form \eqn{laxpoisson}, with the $r$ and $s$ matrices given by Eqs. \eqn{rmat}, \eqn{smat}, we can repeat the analysis performed in \cite{Rajeev:1996kk}, the sole formal difference being in the parameter $ \tau$ of the ref. \cite{Rajeev:1996kk}, which is here replaced by $\bar\tau$. Therefore, adapting the results of \cite{Rajeev:1996kk}, one computes the algebra of monodromy matrices, obtaining 
\begin{equation} \label{poiMgen}
\begin{aligned}
\left\{\mathcal{M}_1(x, y ; \zeta), \mathcal{M}_2 \left(x, y ; \xi\right)\right\}=& \big[r(\zeta, \xi), \mathcal{M}_1(x, y ; \zeta) \mathcal{M}_2\left(x, y ; \xi\right)\big]\\ & +\mathcal{M}_1(x, y ; \zeta) s(\zeta, \xi) \mathcal{M}_2\left(x, y ; \xi\right) \\ & -\mathcal{M}_2\left(x, y ; \xi\right) s(\zeta, \xi) \mathcal{M}_1(x, y ; \zeta),
\end{aligned}
\end{equation}
where $\mathcal{M}_1=\mathcal{M} \otimes \mathbb{1}$, $\mathcal{M}_2=\mathbb{1} \otimes \mathcal{M}$, with $\mathcal{M}=\mathcal{M}^{ij}e_i \otimes e_j$.

It can be verified by direct calculation that the Jacobi identity for the latter results in   an  equation for the $r$ and $s$ matrices, which coincides with \eqn{modiYB}.
As already noticed, it reduces to the classical Yang-Baxter equation for the matrix $r$ when $s$ is zero. Moreover,  observing that 
\be
\Gamma(\zeta, \xi)=r(\zeta, \xi)+ s(\zeta, \xi), \;\;\;\; -  \Gamma( \xi, \zeta)=r(\zeta, \xi)- s(\zeta, \xi)
\ee 
Eq.  \eqn{modiYB} becomes
\begin{equation}\label{ybaxtergamma}
[\hat{\Gamma}_{12}(\zeta_1,\zeta_2), \hat{\Gamma}_{23}(\zeta_2, \zeta_3)]+[\hat{\Gamma}_{13}(\zeta_1,\zeta_3),\hat{\Gamma}_{23}(\zeta_2,\zeta_3)]-[\hat{\Gamma}_{12}(\zeta_2,\zeta_1),\hat{\Gamma}_{13}(\zeta_1,\zeta_3)]=0.
\end{equation}
As a further check of consistency, it can be directly verified   that $r$ and $s$ defined in  \eqn{rmat}, \eqn{smat} do satisfy \eqn{ybaxtergamma}. 

By virtue of \eqn{consmon} the conserved quantities for any value of the parameters are represented by  the  infinite volume limit of the monodromy matrices. In order to compute their Poisson brackets, one needs to calculate the equal points limit of the brackets \eqn{poiMgen}. This requires a careful regularization. The symmetric limit procedure, illustrated in \cite{Maillet:1985ek} and used in \cite{Rajeev:1996kk}, applies here identically. In the infinite volume limit it was shown to give 
\be\label{rMM}
\{  \mathcal{M}(\zeta),  \mathcal{M}(\xi)\}= [r(\zeta,\xi),  \mathcal{M}_1  \mathcal{M}_2]
\ee
Notice that the latter satisfies Jacobi identity only weakly, namely through the symmetric limit procedure \cite{Rajeev:1996kk}.
Hence,  the conserved quantities $\Tr  \mathcal{M}(\zeta)$ are in involution, it being $\Tr (A\otimes B)= \Tr A\cdot \Tr B$, so that  
\be
\{\Tr \mathcal{M}(\zeta),  \Tr \mathcal{M}(\xi)\}= \Tr \{ \mathcal{M}(\zeta),  \mathcal{M}(\xi)\}
\ee
whichis zero because of \eqn{rMM}. Summarising, it holds
\begin{equation}
\{\text{Tr} \mathcal{M}(\zeta), \text{Tr} \mathcal{M}(\xi)\}=0
\end{equation}
which, being zero, satisfies the Jacobi identity strongly. 

It is worth noticing that in the limit $i\alpha \to 0$, $i\tau \to 0$, 
the matrix pair $r,s$ takes the known form for the original WZW model, and by considering $\rho=0$ one recovers the Principal Chiral Model.

Finally, we note that we have obtained a four-parameter family of non-ultralocal integrable models, if we count, besides the two deformation parameters $\alpha$ and $ \tau$, the  two coupling constants $\lambda$ and $\rho$.

To conclude this section, one can add that,  thanks to the duality prescription in Sec. \ref{secdualmodels}, we also have a four-parameter family of Poisson-Lie dual integrable models having $SB(2,\mathbb{C})$ as target configuration space.

\section{Conclusions and Outlook}
\label{sectconclusions}

We considered an alternative canonical formulation of the WZW model based on a two-parameter deformation of the current algebra introduced in \cite{Bascone:2020dcn} and starting from this we obtained a four-parameter (taking also coupling constants into account) family of non-ultralocal integrable models. Being the current algebra of the alternative formulation homomorphic to the Kac-Moody algebra $\mathfrak{sl}(2,\mathbb{C})(\R)$, it was possible to map the algebra to the direct sum of  two $\mathfrak{su}(2)$ Kac-Moody algebras, by means of a complex linear change of basis.
 In the new  basis it was possible to  show the integrability of the models in the so-called Maillet formalism, which is the one suited for non-ultralocal field theories, relying on the existence of  a couple of matrices, so called $r$ and $s$, satisfying a generalised Yang-Baxter equation. Such matrices are generally dynamical, but not in  this case, where they do not depend on spacetime variables. As a byproduct, one naturally   obtains a four-parameter family of integrable Poisson-Lie dual models on the dual group $SB(2,\mathbb{C})$.

 \vspace{5pt}

%





\end{document}